# Exceptionally high saturation magnetisation in Eu-doped magnetite stabilised by spin-orbit interaction

M. Hussein N. Assadi,[1,*] José Julio Gutiérrez Moreno,[2] Dorian A. H. Hanaor,[3] Hiroshi Katayama-Yoshida[4]

[1] School of Materials Science and Engineering, The University of New South Wales, NSW 2052, Australia.

[2] Department of Computer Applications in Science and Engineering, Barcelona Supercomputing Center (BSC), C/ Jordi Girona 29, 08034 Barcelona, Spain.

[3] Fachgebiet Keramische Werkstoffe, Technische Universität Berlin, 10623 Berlin, Germany.

[4] Center for Spintronics Research Network, Graduate School of Engineering, The University of Tokyo, 7-3-1 Hongo, Bunkyo-ku, Tokyo 113-8656, Japan.

h.assadi.2008@ieee.org

The significance of the spin-orbit interaction is very well known in compounds containing heavier elements such as the rare-earth Eu ion. Here, through density functional calculations, we investigated the effect of the spin-orbit interaction on the magnetic ground state of Eu doped magnetite ($Fe_3O_4$:$Eu_{Fe}$). By examining all possible spin alignments between Eu and magnetite's Fe, we demonstrate that Eu, which is most stable when doped at the tetrahedral site, adapts a spin almost opposite the substituted Fe. Consequently, because of smaller spin cancellation between the cations on the tetrahedral site ($Fe_{Tet}$ and $Eu_{Tet}$) and the cations on the octahedral sites ($Fe_{Oct}$), $Fe_3O_4$:$Eu_{Fe}$ exhibits a maximum saturation magnetisation of 9.451 $\mu_B$/f.u. which is significantly larger than that of undoped magnetite (calculated to be 3.929 $\mu_B$/f.u.). We further show that this large magnetisation persists through additional electron doping. However, additional hole doping, which may unintentionally occur in Fe deficient magnetite, can reduce the magnetisation to values smaller than that of the undoped magnetite. The results presented here can aid in designing highly efficient magnetically recoverable catalysts for which both magnetite and rare earth dopants are common materials.

## Introduction

Magnetite ($Fe_3O_4$) has been the most widely used permanent magnet since around the 9th century BCE, with its discovery and the origin of the word magnet being attributed to Magnes the shepherd.[1] In more recent times, magnetite thin films and nanoparticles, particularly in doped forms, have found a wide range of applications in environmental remediation,[2] spintronics,[3,4] catalysis,[5] drug delivery,[6] and biomedical applications.[7] At room temperature, magnetite has a face-centred cubic (fcc) structure (Figure 1). In this structure, one-third of Fe ions occupy tetrahedral sites ($Fe_{Tet}$) with a +3 oxidation state, while the remaining two-thirds occupy the octahedral site ($Fe_{Oct}$) and are equally likely to be in either +2 or +3 oxidation state. This cationic distribution is commonly referred to as an inverse spinel. In a spinel structure, in contrast, all cations at the tetrahedral sites are in the +2 oxidation state, while all cations in the octahedral sites are in the +3 oxidation state. Despite having a significant net magnetic moment measured at ~ 4.1 $\mu_B$ and magnetic saturation of ~ 93 emu g$^{-1}$ at room temperature,[8] magnetite is in fact a ferrimagnet, meaning that the spin of the tetrahedral Fe ions opposes the spin of the octahedral Fe ions.[9,10] Nonetheless, even with the spin cancellations, as Fe ions are in high spin states ($Fe^{2+}$: $S = 2$, $Fe^{3+}$: $S = 5/2$), the remainder of the magnetic moment is still large.

Often magnetite is doped with other cations to improve its electrical, optical or catalytic functionalities. Among the dopants, rare earth elements (lanthanides) are a popular choice due to their 3+ oxidation state and their unique optical and magnetic properties, which are related to the presence of partially filled f electronic orbitals.[11-28] Given the higher atomic mass and substantially larger atomic radii of the rare earth elements compared to Fe and O, one would wonder what the role of the spin-orbit interaction is in determining the structure and electronic features of the rare earth doped magnetite. Spin-orbit interactions originate from the coupling between an electron's spin and the magnetic field induced by the relative motion of the nucleus with respect to the electron.[29] The strength of the magnetic interaction increases with the mass of the element in a complex manner.[30] Summarily, for valence electrons, the spin-orbit interaction strength scales somewhere between the lower bound of the Landau-Lifshitz scaling at $Z^2$ ($Z$ is the atomic number) and the higher bound of $Z^4$ obtained by the hydrogenic approximation.[31] In any case, the spin-orbit interaction becomes significant in determining the magnetic ground state of the 4d, 5d, and 4f compounds.

In the absence of spin-orbit coupling, the orbital moment (*l*) and the spin (*s*) are independent quantum descriptors of a system. Consequently, a compound's magnetic ground state depends only on the relative orientation of the spin of its magnetic ions with respect to each other. Hence, identifying the magnetic ions as having either spin-up or spin-down suffices in describing the magnetic configuration, as rotating all magnetic moments simultaneously by the same angle does not change the energy. However, when spin-orbit interactions become significant, *l* and *s* cease to be independent, as the spin-orbit interaction contains the term *l* · *s*. In such a case, the spin directions of the magnetic moment should be described with respect to the crystallographic directions. Moreover, the spin-orbit interaction often results in magnetic noncollinearity, where the spins on different crystallographic sites may take an arbitrary orientation with respect to one another, instead of being strictly parallel or antiparallel.[32]

Given the argument above, one would anticipate that the spin-orbit interaction's role is significant in rare-earth-doped magnetite systems. However, this interaction is unlikely to play a dominant role in governing the properties of undoped magnetite. Although some previous studies have discussed the structural, optical and magnetic properties of Eu-doped magnetite[12, 16] processed by different experimental methods, more fundamental studies aiming at deeper insights into the magnetic properties of Eu-doped magnetite based on the analysis of spin-orbit interactions are still lacking. Here, motivated by closing this gap, we examine the structural, magnetic and electronic properties of Eu as a representative of the lanthanides doped magnetite. We demonstrate that Eu doping introduces significant noncollinearity, which greatly enhances the saturation magnetisation in magnetite. Furthermore, we show that additional hole doping destabilises the magnetic coupling in Eu-doped magnetite.

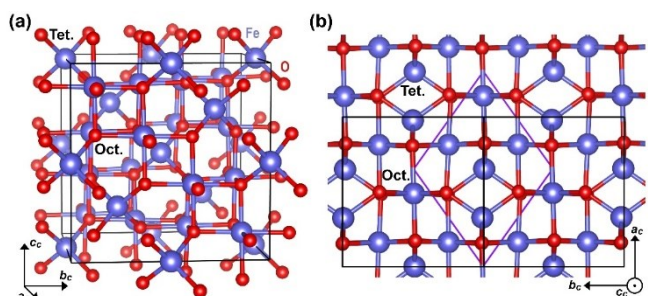

Figure 1. (a) Conventional cell of inverse spinel form $Fe_3O_4$, which is face-centred-cubic, *i.e.*, $Fd\bar{3}m$ (227) space group and containing 56 atoms. Fe ions on the octahedral site have a spin direction opposite to that of Fe ions on the tetrahedral sites. (b) The primitive cell of $Fe_3O_4$ containing 14 atoms is marked with purple boundaries. Light blue and red spheres denote Fe and O, respectively.

**Methods and Models**

Noncollinear density functional calculations, which include the spin-orbit interaction, were performed with VASP,[33, 34] using the projector augmented wave method (PAW) technique[35] and the Perdew–Burke–Ernzerhof (PBE) exchange-correlation functional,[36] which is based on the general gradient approximation. In VASP, the spin-orbit interaction is added to the electronic part of the PAW Hamiltonian and solved variationally, while PAW pseudopotentials account for the mass velocity and Darwin terms. To improve the band description, adequate intra-atomic interaction terms ($U_{eff}$), based on the approach proposed by Liechtenstein *et al.*,[37] were added to Fe 3d and Eu 4f states. The $U_{eff}$ was 5.3 eV ($U$ = 5.3 eV and $J$ = 0.0 eV) for Fe 3d electrons and 7.7 eV ($U$ = 8.7 eV and $J$ = 1.0 eV) for Eu 4f electrons. Comparable values were reported to improve the calculation accuracy of rare-earth compounds[38-40] and magnetite.[41, 42] Furthermore, as shown in Figure S1, the applied $U_{eff}$ values were necessary to adequately localise the 4f and 3d electrons. The energy cut-off was set at 520 eV. The precision key for the rest of the parameters was set high. The noncollinear calculations were initiated with the wave functions calculated with the spin-polarised collinear method to aid convergence. The results for collinear and noncollinear calculations are compared in Figure S2. Accordingly, noncollinear calculations consistently resulted in lower total energies, indicating higher stability.

For the simulation of $Eu_{Fe}$ doping, a tetrahedral Fe in the primitive $Fe_3O_4$ cell, enclosed in purple boundaries in Figure 1b, was substituted with Eu. This primitive cell initially contained two tetrahedral Fe and four octahedral Fe. The substitution of Eu in the tetrahedral site was justified by examining all spin configurations of octahedrally Eu-doped magnetite, both with and without carrier co-doping. As shown in Figure S3, we found that Eu substitution at $Fe_{Oct}$ sites had at least ~ 1 eV higher formation energy than Eu substituting for $Fe_{Tet}$. In other words, Figure S3 indicates that under equilibrium synthesis, Eu energetically favours tetrahedral substitution. In this simulation, the replacement of one Fe for Eu results in a cationic doping concentration of 16.67%. For simulating electron and hole doping, one oxygen was substituted with either N ($N_O$) or F ($F_O$).[43] $N_O$ simulates hole doping, while $F_O$ simulates electron doping at 12.5% carrier doping with respect to the total concentration of O. Adding anionic co-dopants for simulating carrier co-doping is a common method to avoid simulations of charged supercell which introduce artificial dipole-dipole interactions.[44, 45] As articulated in Figure S4, N/F co-dopants were placed at the most stable sites within the lattice.

For geometry optimisation, the internal coordinates and the lattice parameters were allowed to relax to energies and forces smaller than $10^{-6}$ eV and 0.01 eV Å$^{-1}$, respectively A dense k-point mesh, generated with the Monkhorst-Pack scheme of ~ 0.05 Å$^{-1}$ spacing, consisting of 256 irreducible sampling points in the Brillouin zone, was used for geometry optimisation, ensuring accurately calculated forces for obtaining relaxed structures. This level of accuracy was required as final geometries are sensitive to the magnetic configuration.[46] The density of states (DOS) was calculated with a denser mesh of 343 irreducible k-points to adequately capture sharply localised states, along with the tetrahedron smearing method with Blöch corrections at a $\sigma$ of 0.05.

Six initial spin configurations, shown in Figure 2 and labelled (a) through (f), were examined to compare the relative stability of different possible magnetic alignments among Fe ions and the Eu dopant. In configuration (a), $Eu_{Fe}$ and $Fe_{Tet}$ were set to have opposite spins to the $Fe_{Oct}$ ions, similar to the ferrimagnetic configuration of pristine $Fe_3O_4$. In configuration (b), the spin direction of $Eu_{Fe}$ was set opposite to that of $Fe_{Tet}$ and parallel to the spin of the $Fe_{Oct}$ ions. Configurations (a) and (b) were chosen to examine whether or not $Eu_{Tet}$ adopts a spin direction similar to the replaced $Fe_{Tet}$ ion. The spin direction of $Fe_{Tet}$ and $Eu_{Fe}$ in configurations (c) and (d) are similar to those in (a) and (b), except that the spin direction of a $Fe_{Oct}$ ion adjacent to $Eu_{Fe}$ was flipped to examine whether $Eu_{Fe}$ affects the spin direction of any octahedral Fe. In configuration (e), the spin direction of $Eu_{Fe}$ conforms to that of the $Fe_{Tet}$ it replaces, while the spin direction of the remaining $Fe_{Tet}$ was flipped. This configuration served to examine if $Eu_{Tet}$ would affect the spin direction of the remaining tetrahedral Fe. In configuration (f), all cations' spin direction was set parallel to examine the possibility of $Eu_{Te}$ inducing ferromagnetic spin alignment in magnetite.

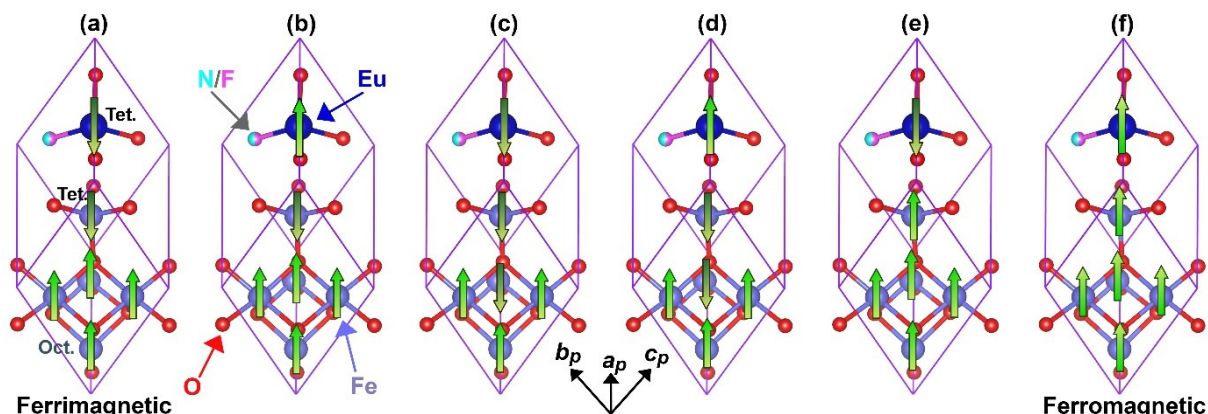

Figure 2. The initial spin alignments investigated for finding the ground state of Eu-doped and carrier co-doped $Fe_3O_4$. Hole or electron doping was simulated by replacing an oxygen ion with a nitrogen or fluorine ion.

## Results and Discussions

For the $Fe_3O_4$:$Eu_{Fe}$ system, as shown in Figure 3, the optimised structure of configuration (b) was found to be the most stable. As shown in Figure 4a, the tetrahedral Fe has an almost antiparallel spin to octahedral Fe ions. The $Fe_{Tet}$ spin's angle to the net spin of the $Fe_{Oct}$ ions was practically complete at 179.56°. However, the Eu ion's spin was, to a good approximation, antiparallel to that of $Fe_{Tet}$, having an obtuse angle of 175.16°. The $Eu_{Fe}$'s spin alignment was opposite to what the original tetrahedral Fe would have been, which is parallel to the other $Fe_{Tet}$. As a consequence of the smaller spin cancellation from the tetrahedral sites, the net magnetisation of the $Fe_3O_4$:$Eu_{Fe}$, calculated to be 9.451 $\mu_B$/f.u., was significantly larger than that of the pristine $Fe_3O_4$, which has been measured to be ~ 4.1 $\mu_B$/f.u.,[8] and calculated to be 3.929 $\mu_B$/f.u. (Table S1).

For the hole-doped $Fe_3O_4$:$Eu_{Fe}N_O$ system, the most stable structure corresponds to the optimised configuration (d). Moreover, as seen in Figure 4b, the optimisation resulted in a significant spin rotation with respect to the initial spin configuration of Figure 2d. Here, resembling the $Fe_3O_4$:$Eu_{Fe}$ structure, Eu's spin alignment was almost opposite that of the $Fe_{Tet}$, taking an obtuse angle of 158.78°. Furthermore, one of the octahedral Fe ions (marked with a star in Figure 4b) was aligned against the rest of the octahedral Fe ions. The angle between this $Fe_{Oct}$'s spin and the rest of $Fe_{Oct}$ ions' net spin was 141.38°. The strong noncollinearity and the substantial spin cancellation among $Fe_{Oct}$ ions resulted in a net magnetisation of 4.929 $\mu_B$/f.u. for the $Fe_3O_4$:$Eu_{Fe}N_O$ system, which is smaller than that of $Fe_3O_4$:$Eu_{Fe}$, but still larger than that of undoped $Fe_3O_4$.

For the electron-doped $Fe_3O_4$:$Eu_{Fe}F_O$ system, the most stable structure was the optimised configuration (b). Here again, the relaxation resulted in a significant spin rotation with respect to the initial spin alignment. Nonetheless, similar to the last two systems, the Eu ion's spin alignment was against that of the tetrahedral Fe, forming an obtuse angle of 154.44°. The spin directions of the $Fe_{Oct}$ ions were nearly parallel and opposing that of the $Fe_{Tet}$. The angle between the net spin on $Fe_{Oct}$ ions and that on the $Fe_{Tet}$ ion was almost complete at 179.96°. The net magnetisation for the $Fe_3O_4$:$Eu_{Fe}F_O$ system was 7.937 $\mu_B$/f.u. which is still more prominent than the undoped $Fe_3O_4$'s magnetisation but smaller than that of the singly Eu-doped system. The reduced magnetisation is mainly because the doped electron reduces one $Fe^{3+}$ ($d^5$) ion to a $Fe^{2+}$ ($d^6$) with a smaller spin number.

The net magnetisations calculated above can provide an upper-bound estimation for the saturation magnetisation in real compounds below the paramagnetic transition temperature. Given the nature of the simulation, which is periodic in three dimensions, the calculated net magnetisations result in a similar magnetic saturation in a perfectly ordered, maximally dense, infinitely large and single-domain specimen under a strong external magnetic field. Any deviation of this ideal case lowers the magnetic saturation. The maximum temperatures at which the calculated net magnetisations would be stable can be quantitively estimated from the stability margin against the second most stable spin configuration. The larger this merging of stability is, the higher thermal fluctuations would be required to destabilise the most stable spin alignment. For example, earlier DFT calculations indicate that in undoped $Fe_3O_4$, the ferrimagnetic state is more stable than the ferromagnetic state (all Fe ions set with parallel spin alignment) by 0.458 eV/f.u.,[47] resulting in an 858 K Curie temperature.[48]

As shown in Figure 3, the next most stable spin configuration for the $Fe_3O_4$:$Eu_{Fe}$ system is the optimised configuration (e), for which the total energy is 0.306 eV/f.u. higher than the most stable configuration (b). This relatively significant energy difference implies room-temperature stability for the optimised configuration (b), although this magnetisation is not as thermally stable as that of undoped $Fe_3O_4$. For the $Fe_3O_4$:$Eu_{Fe}N_O$ system, the total energy of the second most stable configuration, optimised configuration (c), is merely 0.023 eV/f.u. higher than the optimised configuration (d), i.e., the most stable configuration. The total magnetisation of configuration (c) is 1.500 $\mu_B$/f.u. Similarly, for the $Fe_3O_4$:$Eu_{Fe}F_O$ system, the optimised configuration (d) is the second most stable and has a total energy that is 0.024 eV/f.u. higher than that of the most stable configuration (b). In this case, the optimised configuration (d) has a total magnetisation of 6.081 $\mu_B$/f.u. Given the small margin of stability of the magnetic ground state of the hole- and electron co-doped $Fe_3O_4$:$Eu_{Fe}$, one does not anticipate their realisation at room temperature as their small margin of stability against competing magnetic configurations may not prevail against thermal fluctuations. Moreover, the competing magnetic phase in hole-doped $Fe_3O_4$:$Eu_{Fe}N_O$ has substantially smaller net magnetisation at about the third of the most stable configuration. For the electron-doped $Fe_3O_4$:$Eu_{Fe}O_F$ system, the competing magnetic phase still has a comparable net magnetisation to the most stable configuration.

Eu doped magnetite nanoparticles of ~ 12 nm size, synthesised with coprecipitation technique and calcinated at 260 °C, showed magnetic ordering at room temperature in an earlier experiment.[12] However, the measured magnetic saturation of 23.6 emu g$^{-1}$ was smaller than undoped magnetite nanoparticles prepared with the coprecipitation technique. Based on our results, the smaller magnetic saturation of this experiment can be partly explained by unintentional hole doping, probably via Fe vacancies. In a later experiment on Eu doped magnetite nanoparticles with the morphology of hollow nano-spheres of ~ 300 nm diameters, the magnetic saturation was found to be higher at 60.81 emu g$^{-1}$ at room temperature upon 3.79% Eu doping.[16] Although this trend quantitively agrees with our simulation, a direct comparison is impossible since the latter sample was porous. One should note that high surface area and subsequently higher surface spin disorder in magnetite nanoparticles lead to relatively smaller saturation magnetisation than bulk samples. Consequently, the magnetic saturation of magnetite nanoparticles can be as low as ~ 30 emu g$^{-1}$.[49] Relatively dense bulk magnetite has a saturation magnetism of ~ 93 emu g$^{-1}$ at room temperature, equivalent to a magnetic moment of 3.86 $\mu_B$/f.u. and close to the previously calculated value of ~ 4.1 $\mu_B$/f.u.[50]

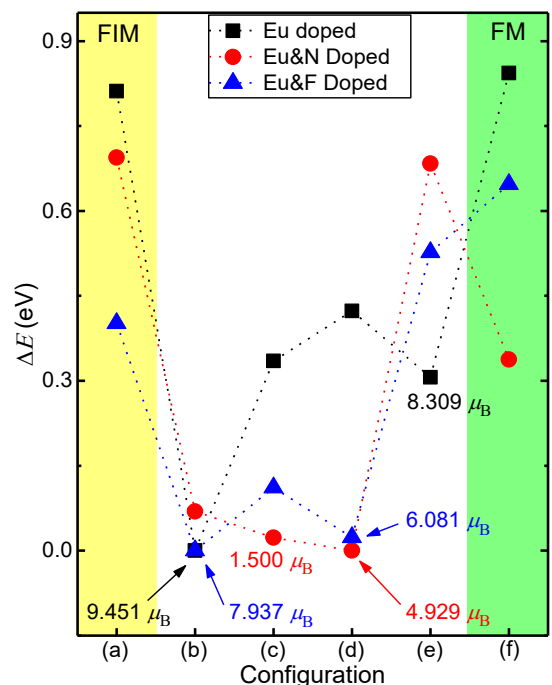

Figure 3. The relative energy of the optimised configurations presented in Figure 2. FIM and FM stand for ferrimagnetic and ferromagnetic, respectively. The net magnetisation of the two most stable configurations is also provided.

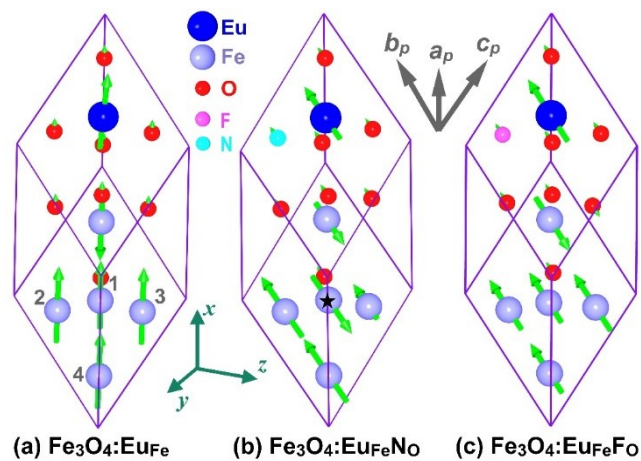

Figure 4. The most stable spin alignments of the (a) $Fe_3O_4$:$Eu_{Fe}$, (b) $Fe_3O_4$:$Eu_{Fe}N_O$ (p-type doping), and (c) $Fe_3O_4$:$Eu_{Fe}F_O$ (n-type doping) based on noncollinear calculations. The primitive cell axes are shown at the top, while the orthogonal frame used for projecting the spinors is shown at the bottom. The x axis is parallel to the $\boldsymbol{a_p} + \boldsymbol{b_p} + \boldsymbol{c_p}$ vector.

Structurally, all optimised systems deviated from the strict cubic symmetry applied at the beginning of the geometry optimisation process, which was based on pristine $Fe_3O_4$ above the Verwey transition point. At temperatures above Verwey transition, the pristine cubic $Fe_3O_4$ unit cell, Figure 1a, has a lattice parameter of 8.394 Å. Our calculated lattice parameter for pristine ferrimagnetic $Fe_3O_4$ was 8.431 Å, which agrees well with the measurements. The cubic $Fd\bar{3}m$ geometry has a trigonal primitive representation, which is shown in Figure 1b. The conventional lattice parameter ($a_c$) is related to the primitive parameter ($a_p$) through $a_p = a_c/\sqrt{2}$. For pristine $Fe_3O_4$, $a_p$ was calculated to be 6.0324 Å. Furthermore, the inclusion of the spin-orbit interaction did not significantly change the calculated lattice parameter relative to collinear calculations (Table S1), indicating the negligible structural effect of the spin-orbit interaction in pristine magnetite. Below Verwey transition temperature of ~ 120 K, however, cubic magnetite undergoes a monoclinic distortion accompanied by a drop in conductivity by two orders of magnitude, which is attributed to the localisation of electrons on two distinct octahedral $Fe^{2+}$ and $Fe^{3+}$ species.[9,10,51,52] Similarly, For $Eu_{Fe}$, $Eu_{Fe}N_O$, and $Eu_{Fe}F_O$ doped systems, the relaxed primitive parameters, subscripted with $p$ and shown in Table 1, cannot be converted to a conventional cubic system. However, rigorous symmetry detection, with a tolerance of $10^{-4}$ Å equal to the calculations' accuracy, could relate these primitive cells to less symmetric conventional cells, of which the parameters indicated by subscript $c$, are also presented in Table 1. Our calculations show that the substitution of Fe by large Eu cations affects the crystal structure and expands the lattice parameters, in line with recent experimental observations in Eu-doped hematite.[53]

Table 1. The relaxed primitive lattice parameters ($a_p$) of the doped systems and their corresponding conventional lattice parameters ($a_c$), found through a symmetry detecting algorithm. Lattice parameters ($a, b, c$) are given in Å and angles ($\alpha, \beta, \gamma$) in °. Based on DFT calculations, for pristine $Fe_3O_4$, $a_p = b_p = c_p = 6.0324$ Å and $\alpha_p = \beta_p = \gamma_p = 60°$.

|  | $Fe_3O_4$:$Eu_{Fe}$ | $Fe_3O_4$:$Eu_{Fe}N_O$ | $Fe_3O_4$:$Eu_{Fe}F_O$ |
|---|---|---|---|
| $a_p$ | 6.214 | 6.040 | 6.098 |
| $b_p$ | 6.214 | 6.040 | 6.098 |
| $c_p$ | 6.214 | 6.201 | 6.098 |
| $\alpha_p$ | 60.00 | 59.09 | 60.00 |
| $\beta_p$ | 60.00 | 59.09 | 60.00 |
| $\gamma_p$ | 60.00 | 62.41 | 60.00 |
| $a_c$ | 6.214 | 10.329 | 6.098 |
| $b_c$ | 6.214 | 6.259 | 6.098 |
| $c_c$ | 15.221 | 8.262 | 14.937 |
| $\alpha_c$ | 90.00 | 90.00 | 90.00 |
| $\beta_c$ | 90.00 | 143.10 | 90.00 |
| $\gamma_c$ | 120.00 | 90.00 | 120.00 |
| Group name | $R3m$ | $C1m1$ | $R3$ |
| Group Number | 160 | 8 | 146 |

Figure 5 shows the partial DOS of the most stable spin configurations, projected on the $x$, $y$, and $z$ axes (frame shown in Figure 4). One noticeable feature in all three compounds is the empty Eu 4f states, marked with blue circles in Figure 5a, d, and g, which have a parallel spin direction to the filled states, demonstrating that $Eu_{Fe}$ takes +3 oxidation state. In $Eu^{3+}$, only 6 electrons occupy the 4f orbitals while the seventh orbital remains empty. Since in the undoped $Fe_3O_4$, tetrahedral sites were occupied by $Fe^{3+}$, one would anticipate that incorporating $Eu^{3+}$ at the tetrahedral site would not change the equal distribution of $Fe^{2+}$ and $Fe^{3+}$ ions in the octahedral sites. However, as we see in the following arguments, this is not the case. As a result, the charge distribution among the Fe ions in the Eu-doped $Fe_3O_4$ does not conform to the simple designation of spinel or inverse spinel.

In the $Fe_3O_4$:$Eu_{Fe}$ system, as shown in Figure 5a, the spin projection on the $x$ axis has the most considerable magnitude, especially for the Fe ions. For Eu, the spin projection on the $y$ and $z$ axes is about one-tenth of the magnitude over the $x$ axis, indicating a slightly larger noncollinearity in the Eu ion, in agreement with the net spin directions of Figure 4a. Moreover, similar to the case of the undoped $Fe_3O_4$, all Fe ions are in a high-spin state. However, the remaining tetrahedral Fe has an oxidation state of +2. For this tetrahedral Fe, the +2 oxidation state means fully occupied minority spin $e$ and $t_2$ states and a singly occupied majority spin $e$ orbital (marked with a green arrow in Figure 5a). Consequently, an octahedral $Fe^{2+}$ should be oxidised to $Fe^{3+}$ to compensate for the tetrahedral $Fe^{2+}$ and maintain charge neutrality. As shown in Table 2, based on net cationic magnetic moments, three out of the four tetrahedral Fe ions are in +3 oxidation state, having fully occupied majority spin $t_{2g}$ and $e_g$ states and completely empty minority spin $t_{2g}$ and $e_g$. The fourth octahedral Fe remains in a +2 oxidation state. For the latter $Fe_{Oct}$ ion, the last electron occupies the minority spin channel of $t_{2g}$ states (marked with a grey arrow in Figure 5a), separated by ~ 9 eV from the majority spin states because of the wider crystal field splitting in octahedral coordination—the crystal field for the tetrahedral $Fe^{2+}$ is ~ 6 eV.

For the hole co-doped $Fe_3O_4$:$Eu_{Fe}N_O$ system, shown in Figure 5d, e, and f, the hole generated by N doping oxidises an octahedral $Fe^{2+}$ to $Fe^{3+}$, leaving no $Fe^{2+}$ on octahedral sites. As a result, all octahedral $Fe^{3+}$ ions have fully occupied majority spin $t_{2g}$ and $e_g$ states and empty minority spin $t_{2g}$ and $e_g$ states. Furthermore, for both occupied and empty $t_{2g}$ and $e_g$ states in the $x$ axis projection, some peaks oppose the majority spin states (marked with grey circles in Figure 5d). These peaks belong to the $Fe_{Oct}(1)$, as marked with a star in Figure 3, with opposing spin to the rest of $Fe_{Oct}$ ions. The tetrahedral $Fe^{2+}$, however, is not oxidised by hole doping, as one of its majority-spin $e$ ortibals remains occupied (marked with green arrows in Figure 5d, e, and f).

For the electron-doped $Fe_3O_4$:$Eu_{Fe}F_O$ system, unlike the previous two systems, the tetrahedral Fe is in a +3 oxidation state, just like the undoped $Fe_3O_4$. The fully occupied spin down $e$ and $t_2$ states and entirely empty majority spin $e$ and $t_2$ states verify the +3 oxidation state of the $Fe_{Tet}$. Having all the ions that occupy the tetrahedral sites, both $Fe_{Tet}$ and $Eu_{Fe}$, at 3+ oxidations state implies that at least two of the octahedral Fe ions should be at +2 oxidation state. Moreover, the electron introduced by $F_O$ doping also must reduce an additional $Fe_{Oct}$. Consequently, as shown in Table 2, three out of four $Fe_{Oct}$ ions in the supercell are in a +2 oxidation state. The additional electron in all $Fe^{2+}$ ions occupies the minority spin $t_{2g}$ states as marked in grey arrows in Figure 5g, h, and i.

The DOS of the $Fe_3O_4$:$Eu_{Fe}$, $Fe_3O_4$:$Eu_{Fe}N_O$ and $Fe_3O_4$:$Eu_{Fe}F_O$ systems show a bandgap of ~ 1 eV, which is wider than the 0.14 eV experimentally reported for pristine $Fe_3O_4$,[54] below Verwey transition at 115 K.[52,55] This bandgap can be explained based on the empirical clues that accompany the bandgap below Verwey transition. These clues are the lowered crystal symmetry from cubic to uniaxial symmetry and the ordering of the $Fe^{3+}$ and $Fe^{2+}$ ions on the octahedral sites, observed in magnetite below Verwey temperature, that promote the broadening of the bandgap. As seen in Table 2, Eu doping alone or with carrier co-doping lowers the crystal symmetry. Moreover, as shown in Table 1, an obvious degree of charge disproportionation can be seen among Fe ions in the Eu-doped and carrier co-doped $Fe_3O_4$ systems. Corroborating our results, many earlier DFT calculations that predicted semimetallic conduction for $Fe_3O_4$ were based on either LDA[56] or GGA[57] level of the theory. These calculations also often reported a uniform oxidation state for $Fe_{Oct}$ of ~ +2.5 without acknowledging the charge disproportionation. This inconsistency is due to the inaccurate description of the exchange term, which typically results in the delocalisation of electrons over the system for standard DFT functionals. However, DFT calculations performed with higher-level theory such as DFT+$U$[58] or Hartree-Fock hybrid DFT[59] can effectively localise d and f electrons in strongly correlated systems, which predicts a charge disproportionation among Fe$_{Oct}$, showing also a bandgap. However, the width of this bandgap is very sensitive to the functional parameters used, but not the charge disproportionation itself. Some also further hypothesised the existence of a bandgap within the bulk of Fe$_3$O$_4$ even above the Verwey transition temperature, attributing the observed conductivity to small polaron hopping instead of band semimetalicity.[60]

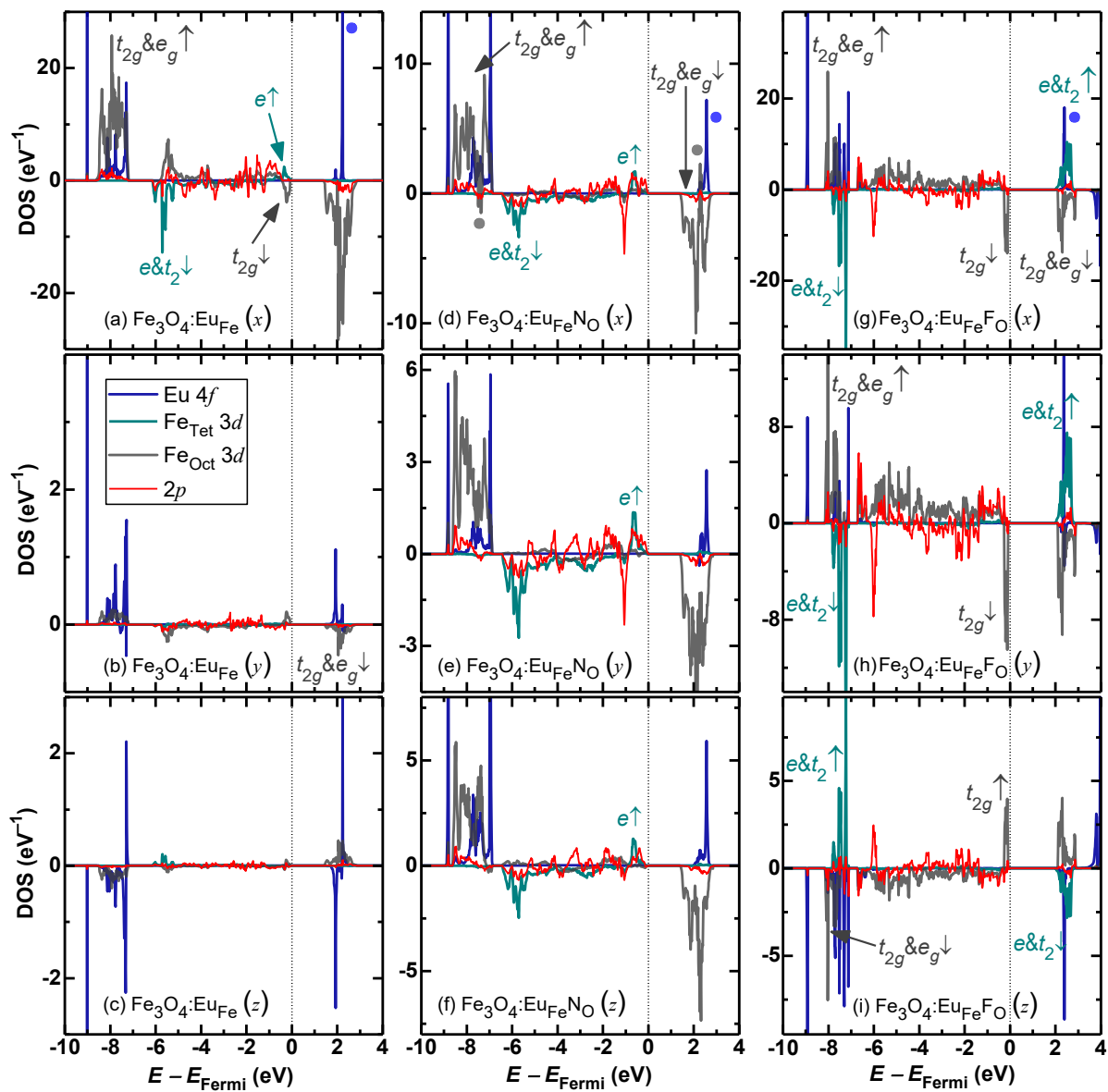

Figure 5. Partial density of states of the Fe$_3$O$_4$:Eu$_{Fe}$ (a, b, and c), Fe$_3$O$_4$:Eu$_{Fe}$N$_O$ (d, e, and f), Fe$_3$O$_4$:Eu$_{Fe}$F$_O$ (g, h, i), projected on the $x$ axis (first row), on $y$ axis (second raw) and on $z$ axis (third raw). The reference frame used for the projection is showed in Figure 4.

Table 2. The calculated localised magnetic moment for Eu and all Fe ions. The location of octahedrally coordinated Fe ions is indexed in Figure 4a. Eu's magnetic moment is smaller than the nominal value because of some degrees of covalency in the Eu-O bond.[47]

| System | Fe$_3$O$_4$:Eu$_{Fe}$ | | Fe$_3$O$_4$:Eu$_{Fe}$N$_O$ | | Fe$_3$O$_4$:Eu$_{Fe}$F$_O$ | |
|---|---|---|---|---|---|---|
| ion | Magnetic moment ($\mu_B$) | Oxidation state | Magnetic moment ($\mu_B$) | Oxidation state | Magnetic moment ($\mu_B$) | Oxidation state |
| Eu$_{Fe}$ | 4.309 | Eu$^{3+}$ | 4.354 | Eu$^{3+}$ | 4.447 | Eu$^{3+}$ |
| Fe$_{Tet}$ | 3.699 | Fe$^{2+}$ | 3.722 | Fe$^{2+}$ | 4.169 | Fe$^{3+}$ |
| Fe$_{Oct}$(1) | 3.805 | Fe$^{2+}$ | 4.275 | Fe$^{3+}$ | 3.744 | Fe$^{2+}$ |
| Fe$_{Oct}$(2) | 4.334 | Fe$^{3+}$ | 4.246 | Fe$^{3+}$ | 3.744 | Fe$^{2+}$ |
| Fe$_{Oct}$(3) | 4.335 | Fe$^{3+}$ | 4.327 | Fe$^{3+}$ | 4.325 | Fe$^{3+}$ |
| Fe$_{Oct}$(4) | 4.337 | Fe$^{3+}$ | 4.277 | Fe$^{3+}$ | 3.744 | Fe$^{2+}$ |

Lanthanide doped magnetite has widely been used as a base or substrate for magnetically recoverable catalysts.[12, 26, 61] Nonetheless, successful recovery requires a strong magnetic response to facilitate magnetic decantation with a permanent magnet.[13, 62] The magnetic attraction between magnetite particles and the permanent magnet is a dipole-dipole interaction proportional to the magnetite's saturation magnetisation.[63] In undoped $Fe_3O_4$, the large magnetisation of ~ 4.1 $\mu_B$ and amply high Curie temperature guarantee ease of magnetic separation at ambient conditions.[8] Given that $Fe_3O_4$:$Eu_{Fe}$ has a magnetisation that is about twice as large as $Fe_3O_4$'s, Eu doping is predicted here to enhance the efficiency of magnetic decantation of catalyst particles. The same statement is also true for the electron co-doped system, $Fe_3O_4$:$Eu_{Fe}F_O$, in which the ground state magnetic phase is not as stable, but competing magnetic phases still produce large magnetisation. However, for hole co-doped $Fe_3O_4$:$Eu_{Fe}N_O$, given the instability of the ground phase and the small magnetisation of the competing phase, one anticipates a drop in the efficiency of magnetic decantation. As a result, one must make sure that in preparing Eu-doped $Fe_3O_4$, no unintentional hole doping occurs. Magnetite tends to be Fe deficient on octahedral sites.[64] These Fe vacancies can be the source of p-type doping in magnetite, which are to be avoided for utilising the full magnetic effect of Eu doping. Lastly, since the noncollinearity is caused by Eu's larger mass, the spin-orbit interaction in Fe compounds containing heavier elements[65-67] would be expected to play a major role, similar to $Fe_3O_4$:$Eu_{Fe}$, and therefore, is quite interesting to investigate.

## Conclusions

We examined the spin alignments among the Fe and Eu ions in Eu-doped magnetite using noncollinear density functional theory at the DFT+$U$ level. We simulated additional carrier co-doping by replacing N for O for producing holes and replacing F for O for producing electrons. We found that Eu dopant preferably substitutes tetrahedral Fe in magnetite, either with or without carrier co-doping. In all cases, $Eu_{Fe}$ was stabilised in a +3 oxidation state, and its spin was at a near-complete angle to the remaining $Fe_{Tet}$, resulting in smaller spin cancellations when compared to the undoped magnetite. This situation results in more substantial saturation magnetisation. The net magnetic moment for $Fe_3O_4$:$Eu_{Te}$, $Fe_3O_4$:$Eu_{Te}N_O$, and $Fe_3O_4$:$Eu_{Te}F_O$, was 9.451 $\mu_B$/f.u., 4.929 $\mu_B$/f.u., and 7.937 $\mu_B$/f.u., respectively. Finally, hole co-doping was found to destabilise the magnetic ground state, lowering the saturation magnetisation and ordering temperature. Electron co-doping, however, was not predicted to significantly decrease the saturation magnetisation.

## Conflicts of interest

There are no conflicts to declare.

## Acknowledgements

The authors gratefully acknowledge the funding of this project by computing time provided by the Paderborn Center for Parallel Computing (PC²). JJGM acknowledges the support from the FusionCAT project (001-P-001722) co-financed by the European Union Regional Development Fund within the framework of the ERDF Operational Program of Catalonia 2014-2020 with a grant of 50% of total cost eligible.

**Electronic Supplementary Information**

# Exceptionally high saturation magnetisation in Eu-doped magnetite stabilised by spin-orbit interaction

M. Hussein N. Assadi,[1,*] José Julio Gutiérrez Moreno,[2] Dorian A. H. Hanaor,[3] Hiroshi Katayama-Yoshida[4]

[1] School of Materials Science and Engineering, The University of New South Wales, NSW 2052, Australia.
h.assadi.2008@ieee.org

[2] Department of Computer Applications in Science and Engineering, Barcelona Supercomputing Center (BSC), C/ Jordi Girona 31, 08034 Barcelona, Spain.

[3] Fachgebiet Keramische Werkstoffe, Technische Universität Berlin, 10623 Berlin, Germany.

[4] Center for Spintronics Research Network, Graduate School of Engineering, The University of Tokyo, 7–3–1 Hongo, Bunkyo-ku, Tokyo 113–8656, Japan.

S2

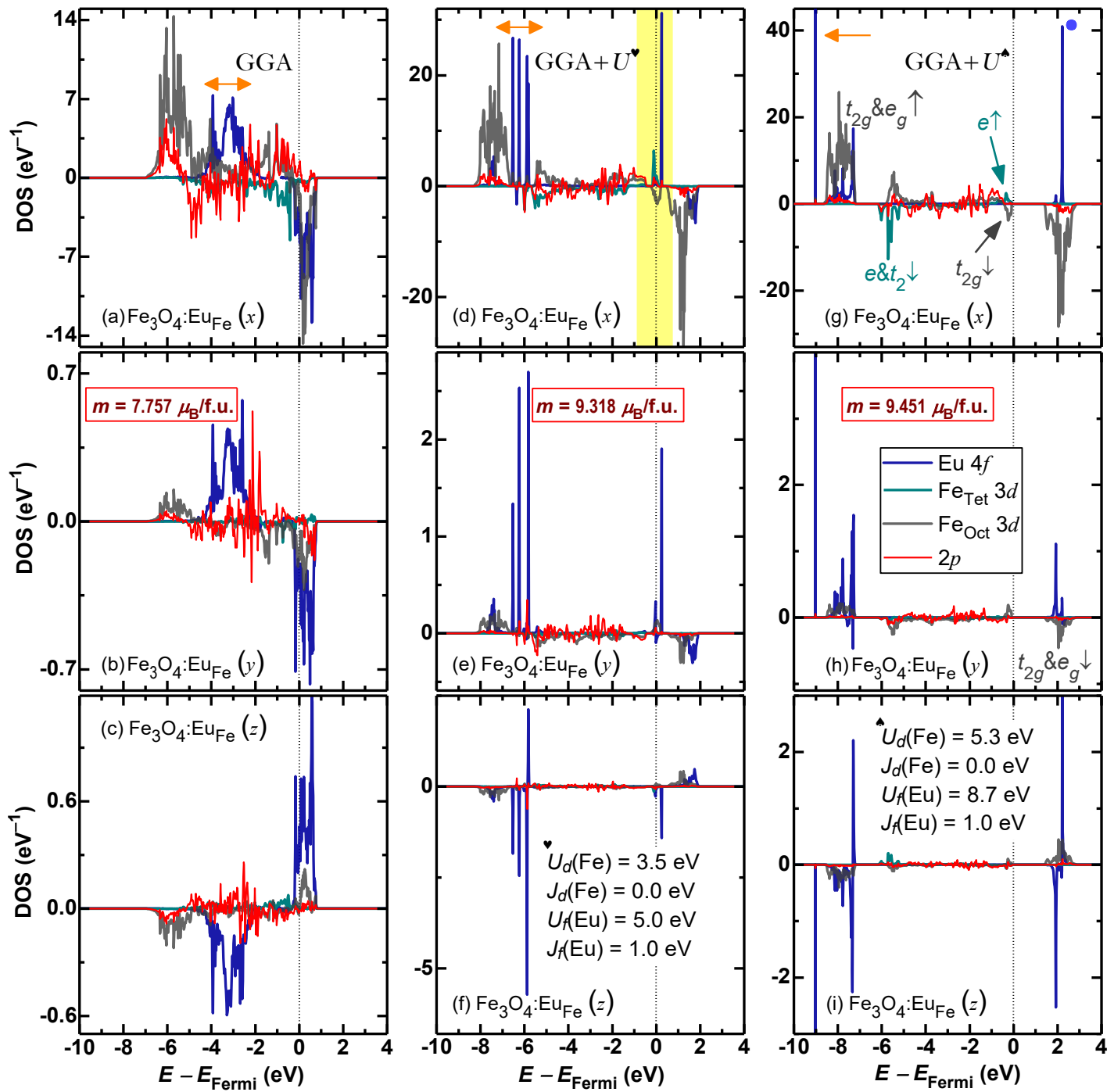

Figure S1. The partial density of states of Fe$_3$O$_4$ doped with tetrahedral Eu at the most stable configuration [configuration (b) of Figure 2] with spin-orbit interaction (SOI) considered. The first column [(a), (b), and (c)] corresponds to the calculation with GGA functional only. The second column [(d), (e), and (f)] corresponds to calculations with relatively smaller $U_{eff}$ values; $U_{eff}$(Eu) = 4 eV, and $U_{eff}$(Fe) = 3.5 eV. The third column [(g), (h), and (i)] corresponds to calculations performed with larger $U_{eff}$ values of 7.7 eV for Eu 4$f$ electrons and 5.3 eV for Fe 3$d$ electrons, reported in the article. The GGA calculations predict a metallic ground state and delocalise the 4$f$ electron, marked with the orange arrow. The GGA + $U$ implementation with the smaller $U_{eff}$ values, although results in a slightly smaller spread of 4$f$ states, still suffers from considerable delocalisation of the 4$f$ states. Moreover, smaller $U_{eff}$ values still predict a metallic state for the compound [shaded area in (d)]. Calculations with the larger $U_{eff}$ values result in a sharp localisation of both filled 4$f$ states [orange arrow in (g)] and empty 4$f$ states (blue circle). The latter band description conforms with the sharp localisation of spacially confined 4$f$ wavefunctions. The bandgap predicted by larger $U_{eff}$ values also conforms with the Mott insulator nature of magnetite at low temperatures. The magnetisation per unit formula is also given for each simulation. The GGA method underestimates the total magnetisation.

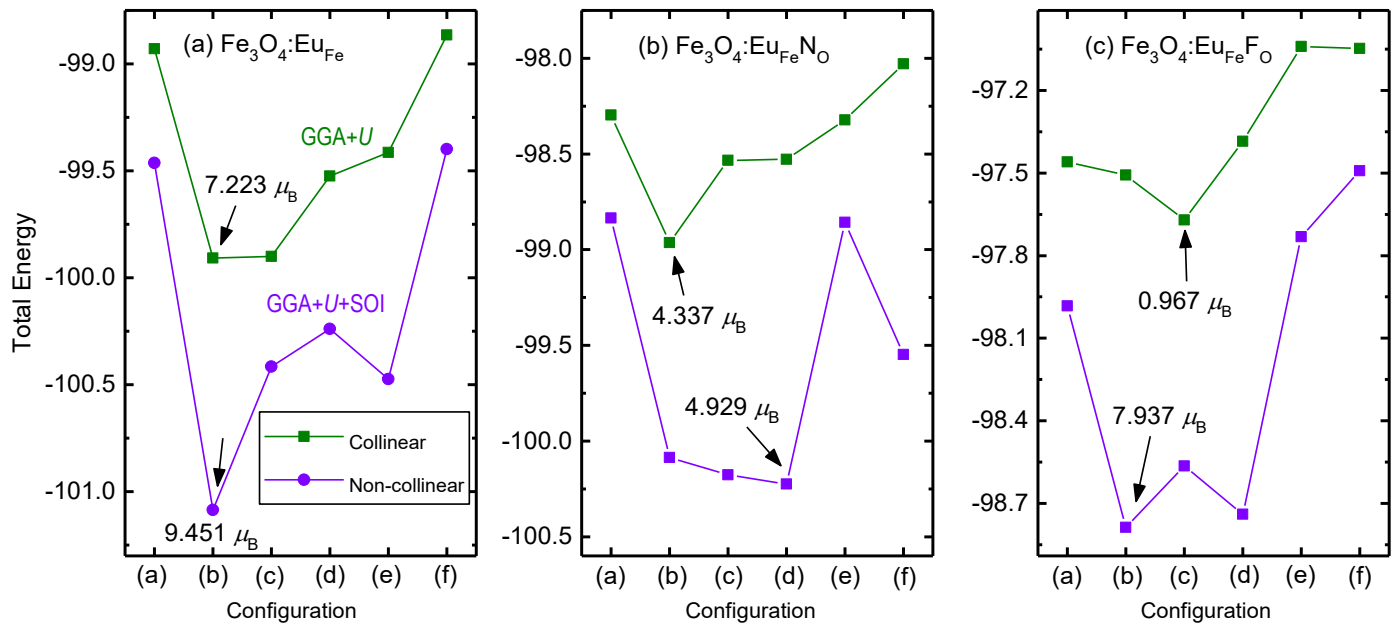

**Figure S2**. The total density functional energies for configuration (a)–(f) of Figure 2 with (purple) and without (green) spin-orbit interaction. The values with spin-orbit interactions were reported in Figure 3 with GGA + $U$ + SOI formalism. The $U_{eff}$ values were 5.3 eV for Fe 3$d$ electrons and 7.7 eV for Eu 4$f$ electrons. The inclusion of the spin-orbit interaction lowers the total energy by about ~1 eV for the most stable configurations. Furthermore, the inclusion of the spin-orbit interaction enhances the total magnetisation of the compounds. SOI calculations, on average, consume 10 to 15 times more computational resources than non-SOI calculations.

**Table S1**. The lattice parameters, compound's total magnetisation and the density functional total energy of the pristine ferrimagnetic $Fe_3O_4$ in primitive unitcell, calculated without and with the spin-orbit interaction. In pristine $Fe_3O_4$, SOI plays a very minor role as it changes the lattice parameters by 0.006%, the total magnetisation by 0.01% and the total energy by 0.02%. The minor contribution of SOI in pristine $Fe_3O_4$ was anticipated as the compound contains only lighter elements.

|  | GGA + $U$ | GGA + $U$ + SOI |
|---|---|---|
| $a_p$ (Å) | 6.0320 | 6.0324 |
| $b_p$ (Å) | 6.0320 | 6.0324 |
| $c_p$ (Å) | 6.0320 | 6.0324 |
| $\alpha_p$ (Å) | 60.000 | 60.000 |
| $\beta_p$ (Å) | 60.000 | 60.000 |
| $\gamma_p$ (Å) | 60.000 | 60.000 |
| $m$ ($\mu_B$/f.u.) | 3.933 | 3.929 |
| $E^t$ (eV) | −93.5395 | −93.5170 |

S4

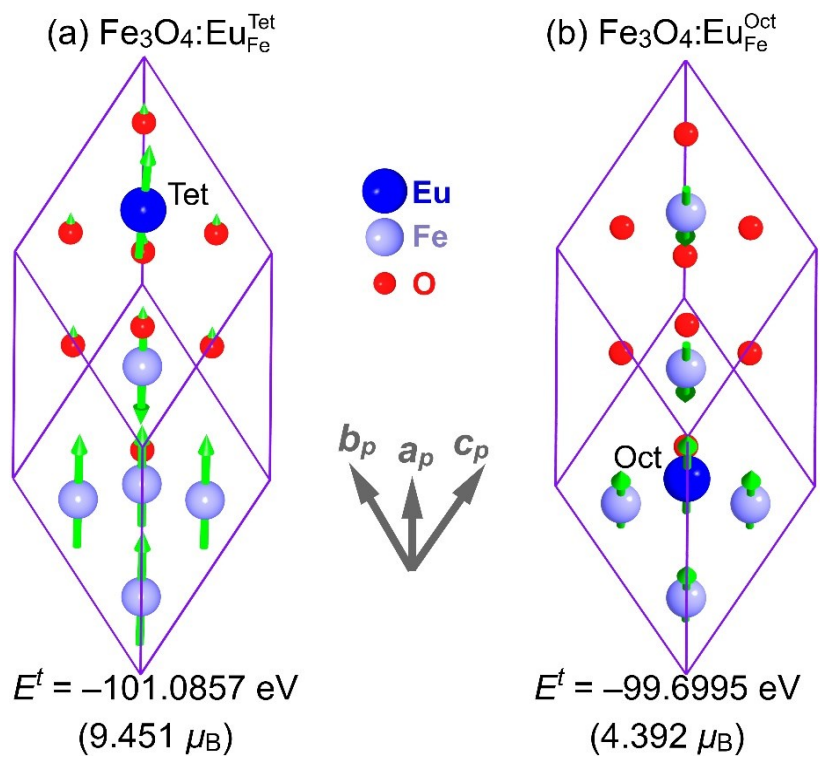

Figure S3. The schematic structures, spin alignments, density functional total energy ($E^t$), and the total magnetisation of Fe$_3$O$_4$ doped with tetrahedral Eu (a) and octahedral Eu (b). Both structures correspond to the most stable spin configurations obtained with the GGA + $U$ + SOI formalism. The tetrahedral Eu doping was more stable by 1.3862 eV.

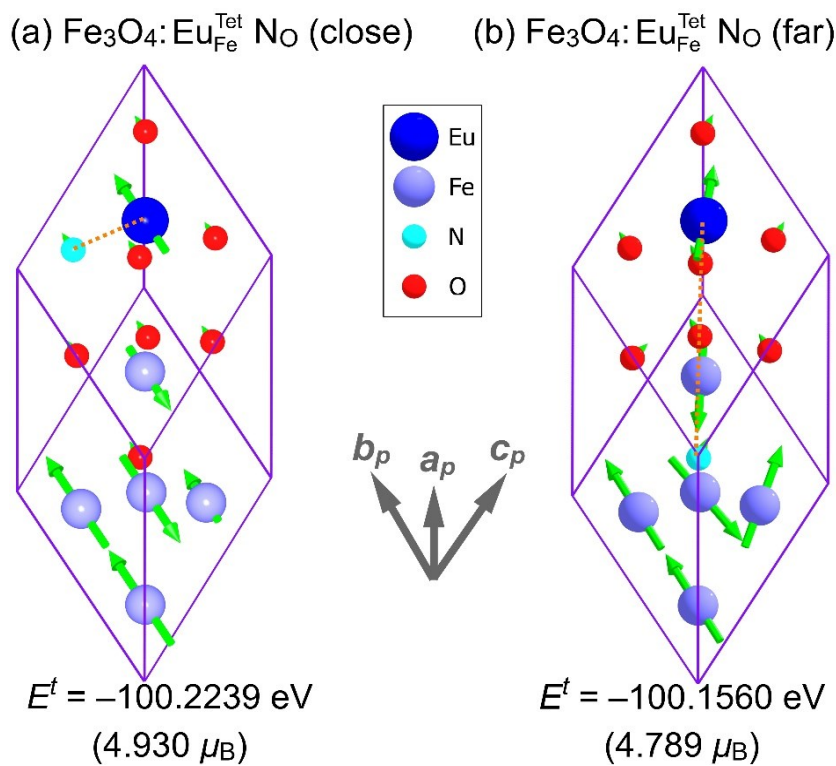

Figure S4. The effect of anionic co-dopant's placement on the total energy and saturation magnetisation is examined here for the case of N codopant simulating hole doping. Among two possibilities for N placements, the site coordinating Eu in (a), i.e., the closest site to Eu, was slightly more stable than the site further away from Eu in (b).